\documentclass{article}
\usepackage{spconf,amsmath,graphicx}

\usepackage{mathrsfs}
\newcommand{\ve}[1]{\textbf{#1}}		
\newcommand{\di}[1]{\mathcal{#1}}		

\usepackage{color}

\title{Language and Noise Transfer in Speech Enhancement Generative Adversarial Network}
\name{Santiago Pascual$^1$, Maruchan Park$^2$, Joan Serr\`a$^3$, Antonio Bonafonte$^1$, Kang-Hun Ahn$^2$\thanks{M.~Park and K.-H.~Ahn were supported by the National Research Foundation of Korea (NRF) grant funded by the Korea goverment (MSIP), NRF-2017R1A2B3010002. S.~Pascual and A.~Bonafonte were supported by the project TEC2015-69266-P (MINECO/FEDER, UE). We thank Cassia Valentini for providing the noise data and the mixing scripts. 
}}

\address{
$^1$ Universitat Polit\`ecnica de Catalunya, Barcelona, Spain \\
$^2$ Chungnam National University, Daejeon, Republic of Korea \\
$^3$ Telef\'onica Research, Barcelona, Spain
}
\begin{document}
%
\maketitle
\begin{abstract}

Speech enhancement deep learning systems usually require large amounts of training data to operate in broad conditions or real applications. This makes the adaptability of those systems into new, low resource environments an important topic. In this work, we present the results of adapting a speech enhancement generative adversarial network by fine-tuning the generator with small amounts of data. We investigate the minimum requirements to obtain a stable behavior in terms of several objective metrics in two very different languages: Catalan and Korean. We also study the variability of test performance to unseen noise as a function of the amount of different types of noise available for training. Results show that adapting a pre-trained English model with 10\,min of data already achieves a comparable performance to having two orders of magnitude more data. They also demonstrate the relative stability in test performance with respect to the number of training noise types.

\end{abstract}
\section{Introduction}
\label{sec:intro}

Speech technologies are at the core of many day-to-day devices, from phones to computers, enabling interaction and communication between devices and people using such devices. 
Speech enhancement, either applied to human or machine speech, is a key technology to facilitate interaction and communication. 
With the goal to improve the intelligibility and quality of speech contaminated by noise~\cite{Loizou2013Book}, speech enhancement is deployed in a myriad of speech processing applications. It can be directly applied as a preprocessing stage in speech recognition and speaker identification systems~\cite{yu2008,maas2012,ortega96icslp} or, in the speech synthesis case, recorded training data can be enhanced to improve synthesis quality~\cite{valentiniinvestigating}. 

In previous work, we proposed an end-to-end speech enhancement system~\cite{pascual2017segan} based on a generative adversarial network~\cite{goodfellow2014generative} (GAN), namely speech enhancement generative adversarial network (SEGAN). SEGAN was proposed in the pursuit of end-to-end speech processing, where signal is enhanced at the raw waveform level, with a one-shot, non-recursive structure. It showed the applicability of latest deep generative modeling to speech enhancement, contrasting with previous techniques that work with spectral features instead of temporal domain. Both objective and subjective results showed its effectiveness as a speech enhancement system.

In the present work, we want to investigate the adaptability of GANs to new languages and noises for the task of speech enhancement. That is, we want to quantify how speech enhancement GANs perform on languages and noises for which they have not been trained for, and assess the amount of new data required to transfer what they have learnt to the new setting. Some of the main research questions we aim to answer are: Does the system perform well for a language it has not been trained for? And for noises? If not, which is the amount of data necessary to adapt the system to the new language/noise? Is it worth to retrain from scratch or is it better to reuse a pre-trained GAN? In both cases, how critical is the selection of training data for adapting the system to the new task? To answer these transfer learning questions in the context of speech enhancement and GANs, we resort to the SEGAN architecture. 



Transfer learning~\cite{pan2010survey} encompasses a number of techniques to adapt pre-trained models towards a personalized task/domain, specially with deep neural networks (DNNs). It has been used in speech recognition to adapt the acoustic models to unseen speakers or mismatched acoustic conditions~\cite{swietojanski2014learning,falavigna2017dnn}. Moreover, transfer learning techniques are applied in speech synthesis to generate speech of different speaker identities with different DNNs and recurrent neural networks~\cite{fan2015multi,pascual2016multi,wu2015study}. In the case of speech enhancement, which is SEGAN's application, it has been shown how DNN-based spectral speech enhancement can be improved by a transfer learning technique, where the top layers are fine tuned for a new language, whilst the lower ones are fixed from a previously trained stage in the original language, where more resources are available~\cite{xu2014cross}. In that work, the authors show how they can already improve their cross-lingual performance with one minute of adaptation speech. Transfer learning has also been devised for model compression of DNN-based spectral enhancement, hence reducing the size of speech enhancement models without performance loss, deepening the network and making it narrower~\cite{wang2017transfer}.

Our contributions are in the exploration of generalization capabilities of SEGAN across languages and noise types, showing that (1) a pre-trained system does not perform well with new languages but, nonetheless, (2) only a few minutes of new training data are needed to drastically improve performance and (3) that a system pre-trained in a different language can save much time to the practitioner (in terms of training data and epochs). In addition, we observe that (4) performance to unseen noises is not affected by the amount of training noises, and that (5) a pre-trained system also helps in obtaining more stable scores. We conclude with the observation that there seems to be two groups of noise types yielding different performances, both in pre-trained and trained-from-scratch systems.

The structure of the paper is as follows. First, SEGAN is introduced. Then, in section~\ref{sec:experimental setup} the experimental details are covered, with the specifics of the architecture and data used to conduct the experiments. Results are discussed in section~\ref{sec:results}, and some conclusions that wrap up our observations from the experiments are discussed in section~\ref{sec:conclusions}.

\section{SPEECH ENHANCEMENT GAN}
\label{sec:segan}

SEGAN~\cite{pascual2017segan} is a deep generative model to denoise speech signals contaminated by additive noise. Considering we have an input noisy signal $\tilde{\ve{x}}$, it learns to map it to the enhanced/clean version $\hat{\ve{x}}$, having as a reference the original clean signal $\ve{x}$. SEGAN follows an adversarial approach during training~\cite{goodfellow2014generative}: a generator (G) network maps $\tilde{\ve{x}}$ into $\hat{\ve{x}}$, and a discriminator (D) network tries to tell the difference between $\hat{\ve{x}}$ and $\ve{x}$, such that G will receive useful clues on what to correct to achieve a realistic enhanced result. It is important to note that this model works in an end-to-end fashion, in the audio sample level (that is, with raw waveforms).

\begin{figure}[t]
\centering
\includegraphics[width=0.5\linewidth]{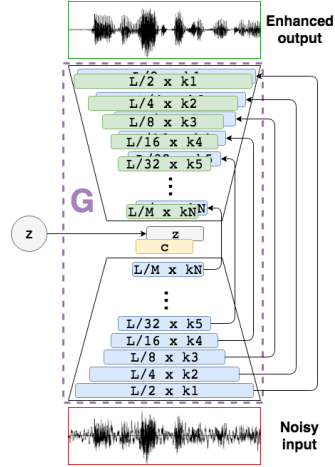}
\caption{\label{fig:segan} SEGAN generator structure. 
}
\end{figure}

The generator network is depicted in Figure~\ref{fig:segan}, where we can see an encoder-decoder fully 1D convolutional structure with skip connections (right side arrows). The encoder is composed of blocks of strided convolutions followed by parametric rectified linear units (PReLUs)~\cite{he2015delving}. We thus get a convolution result out of every $N$ steps of the filter (where $N$ is the stride, see Figure~\ref{fig:segan}). The encoder's input is a waveform chunk $\tilde{\ve{x}}$ of typically 1\,s, which is compressed into the latent representation $\ve{k} = [\ve{c}|\ve{z}]$, where $|$ denotes concatenation operation, $\ve{z}$ is the latent random vector withdrawn from a prior distribution $\di{Z}$ (Figure~\ref{fig:segan}), and $\ve{c}$ is the deterministic code generated by the encoder. 

Once the latent code $\ve{k}$ is computed, the decoder module reconstructs an enhanced version of the speech by reversing the encoding process with fractional-strided transposed convolutions (also named deconvolutions), followed again by PReLUs. In each decoder convolution, the number of channels is doubled, because there is a concatenation of encoder's feature maps with decoder's hidden activations. These concatenations happen because of the so called skip connections, which have the purpose of passing the fine-grained information of the waveform to the decoding stage. Moreover, they offer a better training behavior, as the gradients can flow deeper through the whole structure~\cite{he2016deep}, with easier paths that perform identity functions and do not distort signals. 

\section{EXPERIMENTAL SETUP}
\label{sec:experimental setup}

\subsection{Database details}

We investigate transfer learning from English to Catalan and from English to Korean, as well as the requirements for noise generalization with two experiments for each target language. Experiment~1 trains SEGAN over different speech durations for two baselines: one pre-trained with English (preeng), and the other one based on random parameters (scratch). With this experiment we want to show the amount of speech data required to saturate performance in the transfer learning process. Experiment~2 trains SEGAN over a number of different noise types for both English and random initializations in order to see how generalization to unseen test noises varies with training noise types. Both experiments~1 and~2 consider unseen speakers, sentences, and noise types in the test sets (further details on experiments below).

The English data set consists of a total of 30~speakers (15~male, 15~female), with each speaker recording 400~sentences from the Voice Bank Corpus~\cite{veaux2013}. To train the English baseline, 28~speakers (14~male, 14~female) were chosen, mixed with 40~noise conditions (10~noise types, 4~SNR: 15, 10, 5, and 0\,dB). Therefore, around 10~sentences have each noise condition~\cite{valentiniinvestigating}. The test set contains 2~speakers contaminated by 20~different conditions from 5~types of noise with 4~SNR each (17.5, 12.5, 7.5, and 2.5\,dB).

We made the Korean data set by recording around 20\,minutes for 
12~speakers (6~male, 6~female; Chungnam National University students). Recordings were conducted in a quiet room with the speaker and recorder together, with sentences chosen from the Korean web portal NAVER Open Podium and NAVER Encyclopedia. The Catalan data set consists of recordings from 12~speakers (6 male and 6 female) as part of the FestCat project~\cite{bonafonte2008festcat}. 
Each speaker recorded at least 1h of short paragraphs, which were selected from a set of novels to achieve phonetic and prosodic coverage. The recordings took place in a recording studio. We selected 20\,minutes/speaker to balance with Korean data for experiments. 

To obtain the training data set for Experiment~1, for each language, we added noise in the same way as the English baseline. The training data set for Experiment~2 was made by adding different amounts of noise types (gradually from 1 to 10). The test data set, for each language, considered 20~noise conditions of 5~different noise types and 4~SNRs from the noise conditions of the training data set for 2~speakers.

\subsection{Data arrangement and experiments}
\label{sec:experims}

\textbf{Experiment 1:} Based on English and random initializations, we train SEGAN over a range of speech durations: 24\,s, 1, 2, 4, 10, 20, 50, 100, and 200\,minutes. We train the models several times with different utterance selection as follows: ten times for the 24\,s, 1, and 2~minute durations, and five times for the 4, 10, 20, 50, 100, and 200\,minutes durations. In this way, we can observe the performance differences between the two initialization schemes, with varying speech durations in order to explain transfer learning and data requirements.

\noindent
\textbf{Experiment 2:} Verification of performance with unseen noises is conducted by training over different noise types. After conducting Experiment 1, we found that 20\,minutes duration of training data is enough to saturate performance, so we use this amount. We perform five trainings with each randomly selected noise, increasing one by one the amount of noises until they are ten. Consequently, training is performed 50~times in total. At this time, as in Experiment~1 different utterances are selected for every training, until the training data duration reaches 20 minutes.This allows us to observe the performance differences between the number of training noise types, to include with transfer learning behavior.

\subsection{SEGAN settings and evaluation metrics}

The English model is trained for 86~epochs with RMSprop~\cite{Tieleman12COURSERA} optimizer, with same hyper-parameters as in the original SEGAN work. All Catalan and Korean models are trained with batches of 100 samples for 30 epochs. The amount of epochs was reduced to a third of the original setting for faster experimentation given the large amount of models. Architecture details remain as in~\cite{pascual2017segan}. 
To evaluate the quality of the enhanced speech, we compute the following objective measures. PESQ $[-0.5,4.5]$: perceptual evaluation of speech quality, using the wide-band version recommended in ITU-T P.862.2~\cite{p862}. CSIG $\in[1,5]$: mean opinion score (MOS) prediction of the signal distortion attending only to the speech signal~\cite{hu2008}. CBAK $[1,5]$: MOS prediction of the intrusiveness of background noise~\cite{hu2008}. COVL $[1,5]$: MOS prediction of the overall effect~\cite{hu2008}. SSNR $[0,\infty)$: segmental SNR~\cite{quackenbush}.

\section{RESULTS}
\label{sec:results}

\begin{figure*}[t]
\includegraphics[width=1\linewidth,height=5.5cm]{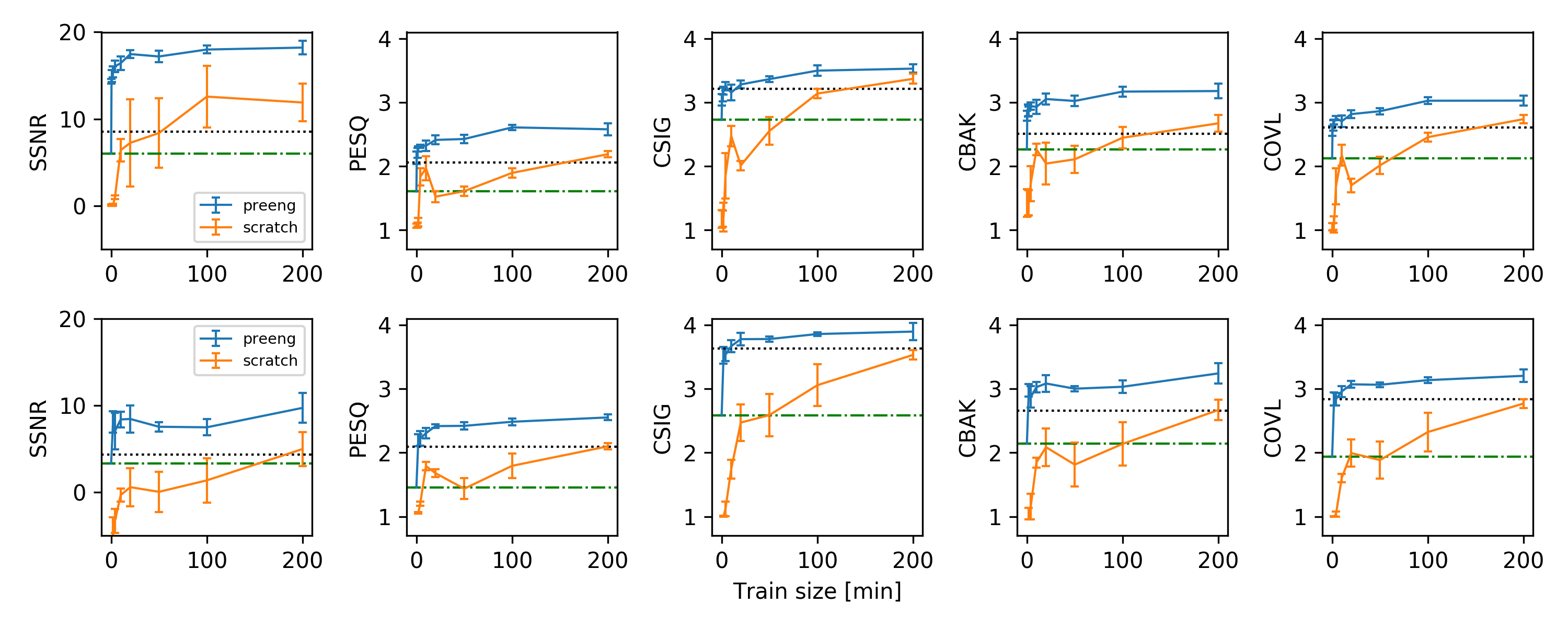}
\caption{\label{fig:results_mixedsnr} Objective metrics for Catalan (top row) and Korean (bottom row). Blue line (preeng): Pre-trained with English. Orange line (scratch): trained from scratch. Green dashed line: SEGAN level without fine tunning. Black dash-dotted line: Noisy level.}
\end{figure*}

The results of the previous metrics as a function training data time for both Catalan and Korean are shown in figure~\ref{fig:results_mixedsnr}. Interestingly, despite the important differences between Catalan and Korean, the results for the two languages show very similar trends. We now elaborate on them.

First of all, we note that the pre-trained English system alone does not perform well with the new languages. We can see it with the green dash-dotted lines, corresponding to the non-adapted SEGAN performance, which is always below the dotted black line, corresponding to the noisy test data without processing. Next, and more importantly, we observe that only a few minutes of new training data are needed to drastically improve performance. In fact, even the minimal training time considered in our methodology (24\,s) significantly improves test performance. To put this result into context, we can compare with the results in~\cite{xu2014cross}, where speech enhancement of a resource-limited language was achieved through cross-language transfer learning of 1\,min. 

Apart from the aforementioned performance increase, we also see that all the evaluation metrics present a knee at about 10\,min, a threshold from which having more training data shows diminishing returns. This is also amazingly small compared to the training time for the English language, which was trained with 9.4\,h and for 86 epochs, thus three times more epochs than the adaptation experiments. These results also indicate that transfer learning can overcome a lack of training epochs with random initialization.

\begin{figure}[!th]
\centering
\includegraphics[width=0.9\linewidth]{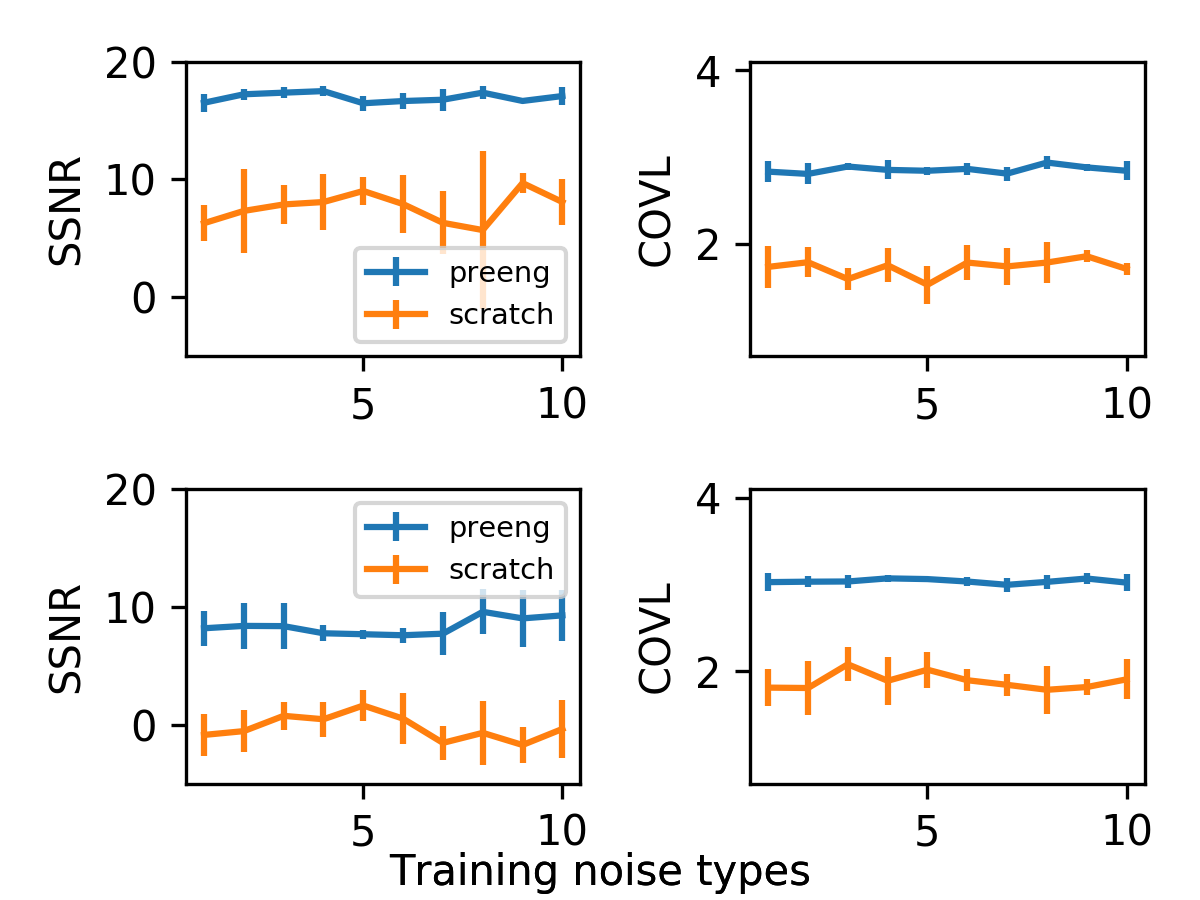}
\caption{\label{fig:nexp} Noise experiment results for Catalan (top row) and Korean (bottom row). Blue lines (preeng): Pre-trained with English. Orange lines (scratch): trained from scratch. Each line shows the resulting objective evaluation of a certain metric depending on the amount of training noises used to test the curve. The amount of noises ranges from 1 to 10, thus the total 10 available in the training set. Qualitatively similar plots were obtained for the PESQ, CSIG, and CBAK metrics.}
\end{figure}

When deploying a system into the real world it is important to consider the final conditions in which the system will work. In speech enhancement, one may be worried about the final noise conditions being quite different from the training ones, and whether the different types of noise in the training data are sufficient to generalize to such unseen noise. To quantitatively assess this situation we conduct Experiment~2 (section~\ref{sec:experims}), whose results are depicted in figure~\ref{fig:nexp}. Strikingly, the objective test metrics do not present a dependence on the number of types of training noise. Whether the training is with English pre-training or from scratch, performance to unseen noises is not affected by the amount of training noises. The only difference between both versions seems to be in the variance of the results, with the scratch version presenting a much larger variance (vertical bars in the orange curves). This implies that, in the case of training from scratch, one should be careful in which types of noise are considered for training. 

\begin{figure}[!th]
\centering
\includegraphics[width=0.95\linewidth]{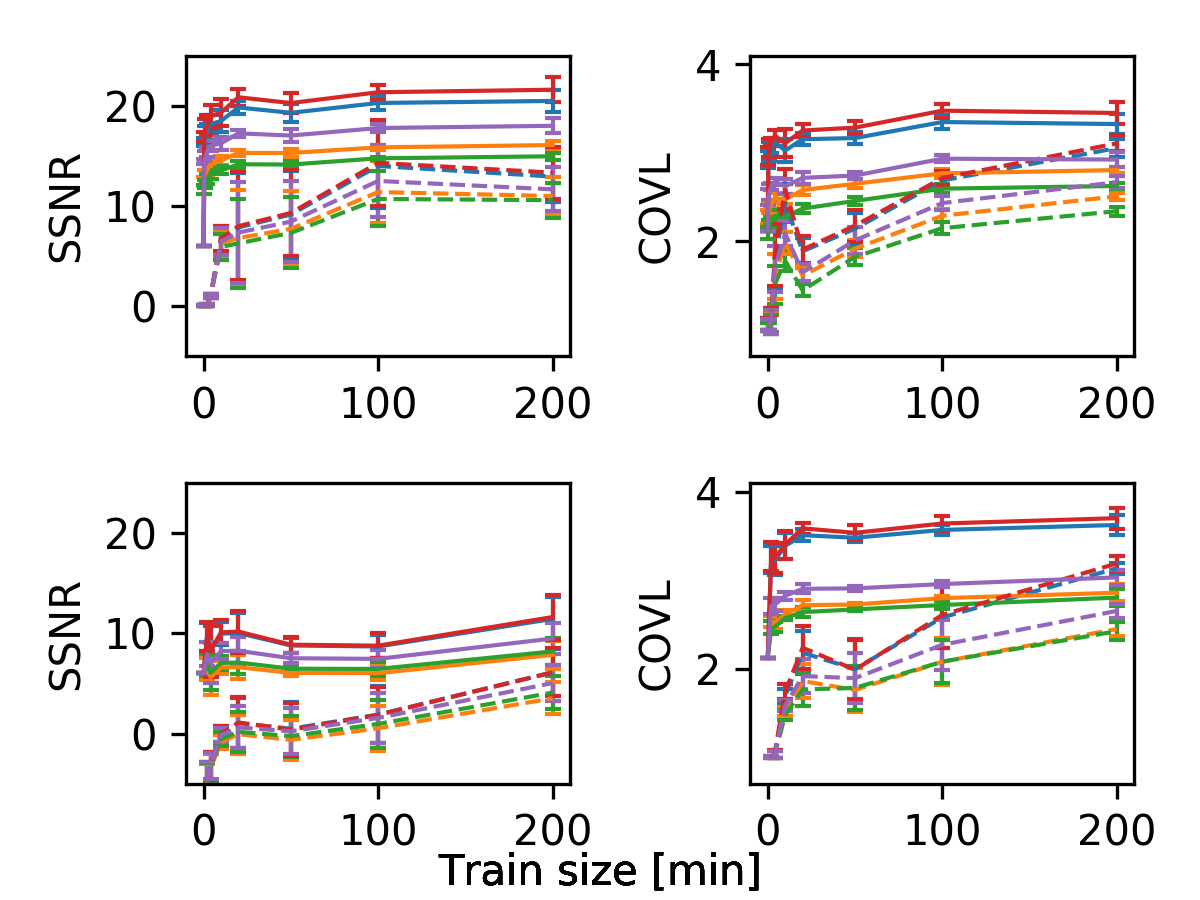}
\caption{\label{fig:2noisetypegroups} Performance on different test noise types. From top to bottom (first solid, then dashed lines), noise types correspond to: office (red), bus (blue), street (purple), living room (orange), and cafe (green) noises. Qualitatively similar plots were obtained for the PESQ, CSIG, and CBAK metrics.}
\end{figure}

Finally, we turn our attention to test noise types (figure~\ref{fig:2noisetypegroups}). Here we also observe consistent behaviors between languages and metrics, with office and bus noise types performing best and street noise, living room, and cafe noises (in this order) performing worse. As a further, more anecdotal note, we see that office and bus noise seem to `cluster' in some metrics (e.g., COVL) in the upper side of the plots, while street noise, living room, and cafe noises also `cluster' in lower metrics.

\section{CONCLUSIONS}
\label{sec:conclusions}

We show that transfer learning is very efficient for inter-language speech enhancement by generative adversarial network. Pre-trained SEGAN with English achieves high performance even for short training time of Catalan and Korean (24\,s) with unseen speakers and noise, driving to adaptability to low resource environments. We also find that the number of noise-type in training is not crucial factor to the performance of the speech enhancement. While training SEGAN is a difficult task, our results imply that one can detour the problem through transfer learning using a pre-trained network.

\vfill
\clearpage

\bibliographystyle{IEEEbib}
\bibliography{strings,refs}

\end{document}